\documentclass[a4,12pt]{article}
\usepackage[dvips]{graphicx,psfrag}
\newcommand{\nc}{\newcommand}
\nc{\bg}{B. Grzadkowski}
\nc{\non}{\nonumber}
\def\dps{\displaystyle}
\def\mib#1{\mbox{\boldmath $#1$}}

\def\bra#1{\langle #1 |} \def\ket#1{|#1\rangle}
\def\vev#1{\langle #1\rangle}

\nc{\barx}{\bar{x}}\nc{\pbarn}{\;\hbox {pb}}
\nc{\fbarn}{\;\hbox {fb}} \nc{\hc}{\hbox {h.c.}}
\nc{\re}{\hbox {Re}} 
\nc{\mev}{\hbox {MeV}} \nc{\gev}{\;\hbox {GeV}}
\def\gesim{\lower0.5ex\hbox{$\:\buildrel >\over\sim\:$}}
\def\lesim{\lower0.5ex\hbox{$\:\buildrel <\over\sim\:$}}
\nc{\prd}[3]{{\it Phys.\ Rev.}\ {{\bf D{#1}} (#2), #3}}
\nc{\prl}[3]{{\it Phys.\ Rev.\ Lett.}\ {{\bf {#1}} (#2), #3}}
\nc{\plb}[3]{{\it Phys.\ Lett.}\ {{\bf B{#1}} (#2), #3}}
\nc{\npb}[3]{{\it Nucl.\ Phys.}\ {{\bf B{#1}} (#2), #3}}
\nc{\ptp}[3]{{\it Prog.\ Theor.\ Phys.}\ {{\bf {#1}} (#2), #3}}
\nc{\zfp}[3]{{\it Z.\ Phys.}\ {{\bf C{#1}} (#2), #3}}
\nc{\epj}[3]{{\it Eur.\ Phys.\ J.}\ {{\bf C{#1}} (#2), #3}}
\nc{\mpla}[3]{{\it Mod.\ Phys.\ Lett.}\ {{\bf A{#1}} (#2), #3}}
\nc{\rmp}[3]{{\it Rev.\ Mod.\ Phys.}\ {{\bf {#1}} (#2), #3}}
\nc{\jhep}[3]{{\it JHEP}\ {{\bf {#1}} (#2), #3}}
\nc{\ijmpa}[3]{{\it Int.\ J.\ Mod.\ Phys.}\
 {{\bf A{#1}} (#2), #3}}
\nc{\ttbar}{t\bar{t}}  \nc{\bbbar}{b\bar{b}}
\nc{\tanb}{\tan \beta}  \nc{\twbdec}{t\to W^+ b}
\nc{\tbwbdec}{\bar{t}\to W^- \bar{b}}
\nc{\epem}{e^+e^-}  \nc{\eett}{\epem \to \ttbar}
\nc{\sigeett}{\sigma_{e\bar{e}\to\ttbar}}
\nc{\wpwm}{W^+W^-}  \nc{\tbar}{\bar{t}}
\nc{\bbar}{\bar{b}}  \nc{\wpp}{W^+}
\nc{\mt}{m_t}  \nc{\mts}{m_t^2}  \nc{\mw}{m_W}  \nc{\mws}{m_W^2}
\nc{\mz}{m_Z}  \nc{\mzs}{m_Z^2}
\nc{\ttbardec}{\ttbar \to W^+W^-\bbbar}
\nc{\wwbb}{W^+W^-\bbbar}  \nc{\sm}{SM}
\nc{\cw}{\cos\theta_W}  \nc{\sw}{\sin\theta_W}
\nc{\sws}{\sin^2\theta_W}  \nc{\sig}{\sigma_{tot}}
\nc{\lp}{{\ell}^+}  \nc{\lm}{{\ell}^-}
\nc{\epsl}{\epsilon_L}  \nc{\cp}{C\!P}
\nc{\gaga}{\gamma\gamma}
\nc{\splus}{s_+}  \nc{\smin}{s_-}  \nc{\eps}{\epsilon}
\nc{\psp}{Ps_+}  \nc{\psm}{Ps_-}  \nc{\lsp}{ls_+}
\nc{\lsm}{ls_-}  \nc{\sss}{s_+s_-}  \nc{\m}{m_t}
\nc{\mq}{m_t^2}  \nc{\mr}{\frac{1}{\m}} \nc{\av}{A_{\gamma}}
\nc{\bv}{B_{\gamma}}  \nc{\az}{A_Z}  \nc{\bz}{B_Z}
\nc{\avs}{A_{\gamma}^2}\nc{\azs}{A_Z^2}  \nc{\bzs}{B_Z^2}
\nc{\dav}{\delta \! A_{\gamma}}  \nc{\dbv}{\delta \! B_{\gamma}}
\nc{\dcv}{\delta C_{\gamma}}  \nc{\ddv}{\delta \! D_{\gamma}}
\nc{\daz}{\delta \! A_Z}  \nc{\dbz}{\delta \! B_Z}
\nc{\dcz}{\delta C_Z}  \nc{\ddz}{\delta \! D_Z}
\nc{\dev}{\delta \! E_{\gamma}}  \nc{\dez}{\delta \! E_Z}
\nc{\dfv}{\delta \! F_{\gamma}}  \nc{\dfz}{\delta \! F_Z}
\nc{\rdav}{{\rm Re}(\delta \! A_{\gamma}) \:}
\nc{\rdbv}{{\rm Re}(\delta \! B_{\gamma}) \:}
\nc{\rdcv}{{\rm Re}(\delta C_{\gamma}) \:}
\nc{\rddv}{{\rm Re}(\delta \! D_{\gamma}) \:}
\nc{\rdaz}{{\rm Re}(\delta \! A_Z) \:}
\nc{\rdbz}{{\rm Re}(\delta \! B_Z) \:}
\nc{\rdcz}{{\rm Re}(\delta C_Z) \:}
\nc{\rddz}{{\rm Re}(\delta \! D_Z) \:}
\nc{\idav}{{\rm Im}(\delta \! A_{\gamma}) \:}
\nc{\idbv}{{\rm Im}(\delta \! B_{\gamma}) \:}
\nc{\idcv}{{\rm Im}(\delta C_{\gamma}) \:}
\nc{\iddv}{{\rm Im}(\delta \! D_{\gamma}) \:}
\nc{\idaz}{{\rm Im}(\delta \! A_Z) \:}
\nc{\idbz}{{\rm Im}(\delta \! B_Z) \:}
\nc{\idcz}{{\rm Im}(\delta C_Z) \:}
\nc{\iddz}{{\rm Im}(\delta \! D_Z) \:}
\nc{\cz}{(1+v_e^2)d\:\!'^2}  \nc{\ci}{v_ed\:\!'}
\nc{\ccz}{v_ed\:\!'^2}  \nc{\cci}{d\:\!'}
\nc{\lspace}{\;\;\;\;\;\;\;\;\;\;}  \nc{\llspace}{\lspace \lspace}
\nc{\beq}{\begin{equation}}  \nc{\eeq}{\end{equation}}
\nc{\bea}{\begin{eqnarray}}  \nc{\eea}{\end{eqnarray}}
\nc{\baa}{\begin{array}}  \nc{\eaa}{\end{array}}
\nc{\bit}{\begin{itemize}}  \nc{\eit}{\end{itemize}}
\nc{\ben}{\begin{enumerate}}  \nc{\een}{\end{enumerate}}
\nc{\bce}{\begin{center}}  \nc{\ece}{\end{center}}
\nc{\ocal}{{\cal O}}
\begin{document}
\pagestyle{empty} \setlength{\footskip}{2.0cm}
\setlength{\oddsidemargin}{0.5cm}
\setlength{\evensidemargin}{0.5cm}
\renewcommand{\thepage}{-- \arabic{page} --}
\def\mib#1{\mbox{\boldmath $#1$}}
\def\bra#1{\langle #1 |}  \def\ket#1{|#1\rangle}
\def\vev#1{\langle #1\rangle} \def\dps{\displaystyle}
\nc{\tb}{\stackrel{{\scriptscriptstyle (-)}}{t}}
\nc{\bb}{\stackrel{{\scriptscriptstyle (-)}}{b}}
\nc{\fb}{\stackrel{{\scriptscriptstyle (-)}}{f}}
\nc{\pp}{\gamma \gamma}
\nc{\pptt}{\pp \to \ttbar}
 \def\thebibliography#1{\centerline{REFERENCES}
 \list{[\arabic{enumi}]}{\settowidth\labelwidth{[#1]}\leftmargin
 \labelwidth\advance\leftmargin\labelsep\usecounter{enumi}}
 \def\newblock{\hskip .11em plus .33em minus -.07em}\sloppy
 \clubpenalty4000\widowpenalty4000\sfcode`\.=1000\relax}\let
 \endthebibliography=\endlist
 \def\sec#1{\addtocounter{section}{1}\section*{\hspace*{-0.72cm}
 \normalsize\bf\arabic{section}.$\;$#1}\vspace*{-0.3cm}}

\vspace{-0.7cm}
\begin{flushright}
$\vcenter{
\hbox{{\footnotesize FUT-07-02}}
{\hbox{{\footnotesize TOKUSHIMA Report}}}
{\hbox{(arXiv:0706.4346)}}
}$
\end{flushright}

\vskip 0.5cm
\begin{center}
{\large\bf Studying possible CP-violating Higgs couplings through}

\vskip 0.15cm
{\large\bf top-quark pair productions at muon colliders}
\end{center}

\vspace{0.2cm}
\begin{center}
\renewcommand{\thefootnote}{\alph{footnote})}
%
Zenr\=o HIOKI$^{\:1),\:}$\footnote{E-mail address:
\tt hioki@ias.tokushima-u.ac.jp},\
Takuya KONISHI$^{\:2),\:}$\footnote{Present affiliation:
{\sl NEC System Technologies, Ltd.}}\ and\
%
Kazumasa OHKUMA$^{\:3),\:}$\footnote{E-mail address:
\tt ohkuma@fukui-ut.ac.jp} 
%
\end{center}

\vspace*{0.2cm}

\vskip 0.2cm
\centerline{\sl $1)$ Institute of Theoretical Physics,\
University of Tokushima}

\centerline{\sl Tokushima 770-8502, Japan}
\vskip 0.2cm
\centerline{\sl $2)$ Graduate School of Human and Natural Environment 
Sciences,}
\centerline{\sl University of Tokushima}
\centerline{\sl Tokushima 770-8502, Japan}

\vskip 0.2cm
\centerline{\sl $3)$ Department of Information Science,\
Fukui University of Technology}
\centerline{\sl Fukui 910-8505, Japan}

\vskip 0.2cm

\vspace*{0.6cm}
\centerline{ABSTRACT}

\vspace*{0.2cm}
\baselineskip=20pt plus 0.1pt minus 0.1pt
We study possible anomalous $C\!P$-violating Higgs couplings to
$\mu\bar{\mu}$ and $t\bar{t}$ fully model-in\-de\-pen\-dent way
through top-quark pair productions at muon colliders. Assuming
additional non-standard neutral Higgs bosons, whose couplings
with top-quark and muon are expressed in the most general
covariant form, we carry out analyses of effects which they
are expected to produce via $C\!P$-violating asymmetries and also
the optimal-observable (OO) procedure under longitudinal and
transverse muon polarizations. We find the measurement of the
asymmetry for longitudinal beam polarization could be useful to
catch some signal of $C\!P$ violation, and an OO analysis might
also be useful if we could reduce the number of unknown parameters
with a help of other experiments and if the size of the parameters
is at least $O(1)\sim O(10)$.

\vfill
PACS:  12.60.Fr, 13.66.Lm, 14.65.Ha, 14.80.Cp

Keywords:
extra Higgs boson, top-quark, muon colliders \\

\newpage
\renewcommand{\thefootnote}{$\sharp$\arabic{footnote}}
\pagestyle{plain} \setcounter{footnote}{0}
\baselineskip=21.0pt plus 0.2pt minus 0.1pt

\sec{Introduction}

It is widely known that the standard model of the electroweak
interaction (SM) has been so far quite successful in describing
various phenomena below the electroweak scale with high precision.
Its top-quark and Higgs-boson sectors are however still not
fully-tested part of the model. If there exists any new physics
beyond the SM within our reach, its effects will be likely to
appear in those sectors. Therefore it is worth to look for
experiments which allow for a comprehensive investigation of
top-quark and Higgs-boson properties.

Anomalous top-quark interactions could be tested, for instance,
at the $e\bar{e}$ colliders in the International Linear Collider
(ILC) project \cite{ILC}. However it is not easy to study Higgs
sector thereby. Muon colliders were proposed as an ideal machine
to explore Higgs properties \cite{First}. From a purely theoretical
point of view, muon colliders are quite similar to $e\bar{e}$
colliders, but the fact that a muon is much heavier than an
electron could provide with a non-negligible difference in
phenomenological studies of Higgs sector.

Indeed many authors have studied how to analyze Higgs-top
interactions at muon colliders. Most of them focused
on the resonance region, i.e., direct Higgs productions, and/or
$\mu\bar{\mu} \to \vev{Higgs} \to t\bar{t}$ in the framework of some
specific models with multi Higgs doublets, like MSSM,
and pointed out that a muon collider will be a useful tool to
identify $C\!P$ properties of Higgs scalars \cite{First}--\cite{GGP}.

As a complementary work to them, we study in this article possible
anomalous Higgs interactions with $\mu\bar{\mu}$ and $t\bar{t}$ in a
fully model-independent way through $\mu\bar{\mu} \to t\bar{t}$
processes. Our main purpose is to clarify to what extent we would
be able to draw a general conclusion on those interactions without
assuming any particular models at muon colliders in off-resonance
region. In other words, we aim to study the possibility and limit
of muon colliders for model-independent analyses of possible new
physics in the top-quark and Higgs-boson sectors.

After describing our calculational framework in section 2, we compute
$C\!P$-violating asymmetries for both longitudinal and transverse beam
polarizations in section 3, where based on the results we also discuss
a detectability of the anomalous-coupling parameters. In section 4,
we study whether the optimal-observable procedure is effective when
we try to determine the anomalous parameters separately. Finally, a
summary is given in section 5.

\sec{Framework}

As mentioned in Introduction, we perform a model-independent analysis
of possible non-standard Higgs interactions with top-quark/muon for
longitudinal and transverse beam polarizations. Let us summarize our
framework first which is the basis of our calculations. Throughout
this paper, we express the standard-model Higgs as $h$ and the
non-standard neutral Higgs as $H$.

\noindent
{\bf Effective amplitude}

The invariant amplitude of $\mu\bar{\mu}\to(\gamma,Z,h,H)\to t\bar{t}$
corresponding to Figure \ref{Feynman} is given as follows:

\begin{figure}[b]
\hspace*{1cm}
\begin{minipage}{14cm}
 \psfrag{t1}{\hspace*{-0.0cm}$t$}
 \psfrag{t2}{\hspace*{-0.0cm}$\bar{t}$}
 \psfrag{m1}{\hspace*{-0.0cm}$\mu$}
 \psfrag{m2}{\hspace*{-0.0cm}$\bar{\mu}$}
 \psfrag{A1}{\hspace*{-0.0cm}$\gamma$}
 \psfrag{A2}{\hspace*{-0.0cm}$Z$}
 \psfrag{A3}{\hspace*{-0.0cm}$h$}
 \psfrag{A4}{\hspace*{-0.2cm}$H$}
\includegraphics[width=13cm,height=5cm]{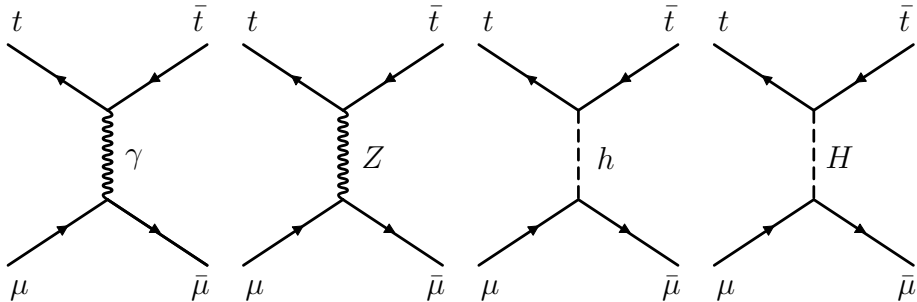}
\caption{Feynman diagrams for $\mu\bar{\mu}\to(\gamma,Z,h,H)
\to t\bar{t}$~~~~~~}\label{Feynman}
\end{minipage}
\end{figure}

\begin{equation}
{\cal M}(\mu\bar{\mu} \to t\bar{t})
= {\cal M}_\gamma + {\cal M}_Z + {\cal M}_h + {\cal M}_H,
\label{Amp}
\end{equation}
where ${\cal M}_{\gamma,Z}$ are the standard $\gamma$ and $Z$ exchange terms
\begin{eqnarray}
&&{\cal M}_\gamma =
D_\gamma(s)\,\bar{u}(\mib{p}_t)
\gamma^\alpha v(\mib{p}_{\bar{t}})
\cdot\bar{v}(\mib{p}_{\bar{\mu}})
\gamma_\alpha u(\mib{p}_\mu),~~~~~~~ \\
&&{\cal M}_Z = D_Z(s)\,\bar{u}(\mib{p}_t)
\gamma^\alpha(A_t+B_t\gamma_5)v(\mib{p}_{\bar{t}})
\nonumber \\
&&\phantom{{\cal M}_Z =}
 \times \bar{v}(\mib{p}_{\bar{\mu}})
\gamma_\alpha(A_\mu + B_\mu \gamma_5)u(\mib{p}_\mu),
\end{eqnarray}
with
\begin{eqnarray}
&&A_t = 1 - (8/3)\sin^2 \theta_W,\ \ B_t = -1,~~~~ \\
&&A_\mu = -1 + 4\sin^2 \theta_W,\ \ \ \ \, B_\mu = 1,
\end{eqnarray}
$D_{\gamma,Z}(s)$ being the propagators multiplied by the coupling constants
\begin{eqnarray}
&&D_\gamma (s)\equiv -\frac{2}{3}e^2 \frac{1}{s}\,, \\ 
&&D_Z(s)\equiv
\frac{g^2}{16\cos^2\theta_W}\frac{1}{s-M_Z^2}\,,
\end{eqnarray}
$e$, $g$, $\theta_W$ being the elementary charge, the $SU(2)$ gauge
coupling, the Weinberg angle respectively, and
$s \equiv (p_\mu + p_{\bar{\mu}})^2$, \\
${\cal M}_h$ is the standard Higgs-boson exchange term
\begin{equation}
{\cal M}_h
=D_h(s)\,\bar{u}(\mib{p}_t)
v(\mib{p}_{\bar{t}})\cdot\bar{v}(\mib{p}_{\bar{\mu}})
u(\mib{p}_\mu),
\end{equation}
while ${\cal M}_H$ is the non-standard Higgs exchange contribution, for
which we assume the most general covariant form:
\begin{eqnarray}
&&{\cal M}_H=D_H(s)\,\bar{u}(\mib{p}_t)(a_t + b_t\gamma_5)
v(\mib{p}_{\bar{t}})
\nonumber \\
&&\phantom{{\cal M}_H=}
\times \bar{v}(\mib{p}_{\bar{\mu}})
(a_\mu + b_\mu\gamma_5) u(\mib{p}_\mu), \label{non-SM}
\end{eqnarray}
with
\begin{equation}
D_i(s) \equiv \frac{m_\mu m_t}{v^2}
\frac{1}{m_i^2 - s - im_i{\mit\Gamma}_i}
\end{equation}
($i=h,\,H$), ${\mit\Gamma}_i$ and $v$ being the total decay width and
the vacuum expectation value of the SM Higgs field. We treat
$a_{t,\mu}$ and $b_{t,\mu}$ as complex numbers to take into account
the possibility that they are form factors.

Readers may claim that we assume only one additional Higgs-boson in spite
of our statement that we perform a fully model-independent analysis.
In fact, our frame can incorporate any number of Higgs exchange terms.
It will be clear by re-expressing such an amplitude as
\begin{eqnarray}
&&{\cal M}[\mbox{\rm\scriptsize Non-SM Higgs}]
=\sum_{i=1}^N D_{H_i}(s)\,\bar{u}(\mib{p}_t)(a_t^i + b_t^i\gamma_5)
v(\mib{p}_{\bar{t}})\cdot
\bar{v}(\mib{p}_{\bar{\mu}})(a_\mu^i + b_\mu^i\gamma_5) u(\mib{p}_\mu)~~
\nonumber \\
&&\phantom{{\cal M}[\mbox{\rm\scriptsize Non-SM Higgs}]}
=\sum_i a_t^i a_{\mu}^i D_{H_i}(s)\,
\bar{u}(\mib{p}_t)v(\mib{p}_{\bar{t}})
\cdot
\bar{v}(\mib{p}_{\bar{\mu}})u(\mib{p}_\mu) \nonumber\\
&&\phantom{{\cal M}[\mbox{\rm\scriptsize Non-SM Higgs}]}
+\sum_i a_t^i b_{\mu}^i D_{H_i}(s)\,
\bar{u}(\mib{p}_t)v(\mib{p}_{\bar{t}})
\cdot
\bar{v}(\mib{p}_{\bar{\mu}})\gamma_5 u(\mib{p}_\mu) \nonumber\\
&&\phantom{{\cal M}[\mbox{\rm\scriptsize Non-SM Higgs}]}
+\sum_i b_t^i a_{\mu}^i D_{H_i}(s)\,
\bar{u}(\mib{p}_t)\gamma_5 v(\mib{p}_{\bar{t}})
\cdot
\bar{v}(\mib{p}_{\bar{\mu}})u(\mib{p}_\mu) \nonumber\\
&&\phantom{{\cal M}[\mbox{\rm\scriptsize Non-SM Higgs}]}
+\sum_i b_t^i b_{\mu}^i D_{H_i}(s)\,
\bar{u}(\mib{p}_t)\gamma_5 v(\mib{p}_{\bar{t}})
\cdot
\bar{v}(\mib{p}_{\bar{\mu}})\gamma_5 u(\mib{p}_\mu).
\end{eqnarray}
Thus, all the contributions can be packed into our parameters as
follows:
\begin{eqnarray}
&&a_t a_\mu = \sum_i a_t^i a_{\mu}^i D_{H_i}(s)/D_{H_1}(s), \\
&&a_t b_\mu = \sum_i a_t^i b_{\mu}^i D_{H_i}(s)/D_{H_1}(s), \\
&&b_t a_\mu = \sum_i b_t^i a_{\mu}^i D_{H_i}(s)/D_{H_1}(s), \\
&&b_t b_\mu = \sum_i b_t^i b_{\mu}^i D_{H_i}(s)/D_{H_1}(s).
\end{eqnarray}

\noindent
{\bf Beam polarization}

The beam polarization, $P$, along one axis (polarization axis) whose direction
is defined by a unit vector $\mib{s}$ is given by
\begin{equation}
P=\frac{\rho_{+s}-\rho_{-s}}{\rho_{+s}+\rho_{-s}},
\end{equation}
where $\rho_{\pm s}$ is the number density of the particle
in each beam whose spin component on this axis is $\pm s$.
We can take into account this polarization by multiplying
the spin vector $s^\alpha$ in the projection operator
$u(\mib{p})\bar{u}(\mib{p})$ and $v(\mib{p})\bar{v}(\mib{p})$
by $P$. That is,
we are to use $(0,P\mib{s})$ as the spin vector in its rest frame.

In the following, we choose the direction of $\mib{p}_{\mu}$ as the $z$ axis
and express the azimuthal angle of $\mib{s}$ as $\phi$. Then the degree of
the longitudinal polarization is given by $P_L=P{s_z}$, that of the
transverse polarization by $P_T=\sqrt{P^2-P_L^2}$, and consequently
\begin{equation}
(0,P\mbox{\boldmath $s$})=(0 , P_T\cos\phi, P_T\sin\phi, P_L).
\label{spin-rest}
\end{equation}
The $\mu$ and $\bar{\mu}$ spin vectors in the $\mu\bar{\mu}$ CM frame are
obtained from (\ref{spin-rest}) via appropriate Lorentz transformations as
\begin{eqnarray}
&&s^\alpha
=(P_L \gamma \beta , P_T \cos\phi , P_T \sin\phi , P_L\gamma)\,,
\label{mu1spin} \\
&&\bar{s}^{\alpha}
=(\bar{P}_L\gamma\beta, \bar{P}_T\cos\bar{\phi},
  \bar{P}_T\sin\bar{\phi}, -\bar{P}_L\gamma)\,,~~~~
\label{mu2spin}
\end{eqnarray}
where
\begin{equation}
\beta\equiv\sqrt{1-4m_\mu^2/s},\quad \gamma\equiv1/\sqrt{1-\beta^2}
\end{equation}
and the momenta of $\mu$ and $\bar{\mu}$ in this frame are
\begin{equation}
p^\alpha=\frac{1}{2}\sqrt{s}(1,0,0,\beta) ,\quad \bar{p}^{\alpha}
=\frac{1}{2}\sqrt{s}(1,0,0,-\beta).    \label{mumom}
\end{equation}

\sec{$\mib{CP}$-violating asymmetries}

It is straightforward to calculate the cross section
$\sigma(\mu\bar{\mu} \to t\bar{t})$ starting from amplitude
(\ref{Amp}) as
\begin{equation}
\frac{d}{d\cos\theta}\sigma(\mu\bar{\mu} \to t\bar{t})
=\frac{1}{32\pi s}\frac{|\mib{p}_t|}{|\mib{p}_{\mu}|}|
{\cal M}(\mu\bar{\mu} \to t\bar{t})|^2.
\end{equation}
We perform this via FORM \cite{FORM}, but the analytical result is a
bit too long to give here explicitly. Therefore we show in the following
our results numerically. Throughout our analysis in this article,
we take $|P_L|=1$ or $|P_T|=1$. It may seem to be an extreme and
unrealistic assumption, but we chose the polarization this way
because our aim here is to know ``to what extent" we could know
about the anomalous interaction (\ref{non-SM}), i.e., we would
like to study the possibility and limit of muon colliders for
model-independent analyses of new physics beyond the standard
model.

\noindent
{\bf Numerical results}

We study two $C\!P$-violating asymmetries $A_L$ and $A_T$, the
former of which is the one for longitudinal beam polarization
\begin{equation}
A_L=\frac{\sigma(++)-\sigma(--)}{\sigma(++)+\sigma(--)},
\end{equation}
and the latter is the one for transverse polarization
\begin{equation}
A_T=\frac{\sigma(\chi=\pi/2)-\sigma(\chi=-\pi/2)}
{\sigma(\chi=\pi/2)+\sigma(\chi=-\pi/2)},
\end{equation}
where $\sigma(\pm \pm)$ express the cross sections for
$P_L=\bar{P}_L=\pm 1$, while $\sigma(\chi=\pm \pi/2)$ are
the ones for $P_T=\bar{P}_T=1$ with $\chi\equiv \phi-\bar{\phi}
=\pm \pi/2$. We here chose $|\chi|$ to be $\pi/2$ since it maximizes
the $C\!P$-violation effects (see, e.g., \cite{GGP}).

Concerning the decay widths of $h$ and $H$, ${\mit\Gamma}_{h,H}$, they
are of course different quantities, but we use the same formula for 
${\mit\Gamma}_H$ as ${\mit\Gamma}_h$ within the standard model
\cite{Hwidth} (see the later discussions). The other SM
parameters are taken as follows:
\[
\sin^2 \theta_W = 0.23,\ \ \ M_Z=91.187\ {\rm GeV},\ \ \
v=246\ {\rm GeV},
\]
\[
m_t=174\ {\rm GeV},\ \ \ m_{\mu}=105.658\ {\rm MeV},\ \ \
m_h=150\ {\rm GeV}.
\]

\begin{figure}[htbp]
\hspace*{1.5cm}
\begin{minipage}{14cm}
 \psfrag{AL}{$A_L$}
 \psfrag{rs}{$\sqrt{s}$}
 \psfrag{G}{GeV}
 \psfrag{mH3}{\begin{footnotesize}$m_H=$300 GeV\end{footnotesize}}
 \psfrag{mH4}{\begin{footnotesize}$m_H=$400 GeV\end{footnotesize}}
 \psfrag{mH5}{\begin{footnotesize}$m_H=$500 GeV\end{footnotesize}}
\rotatebox{270}{\includegraphics[width=9.5cm,height=10.5cm]{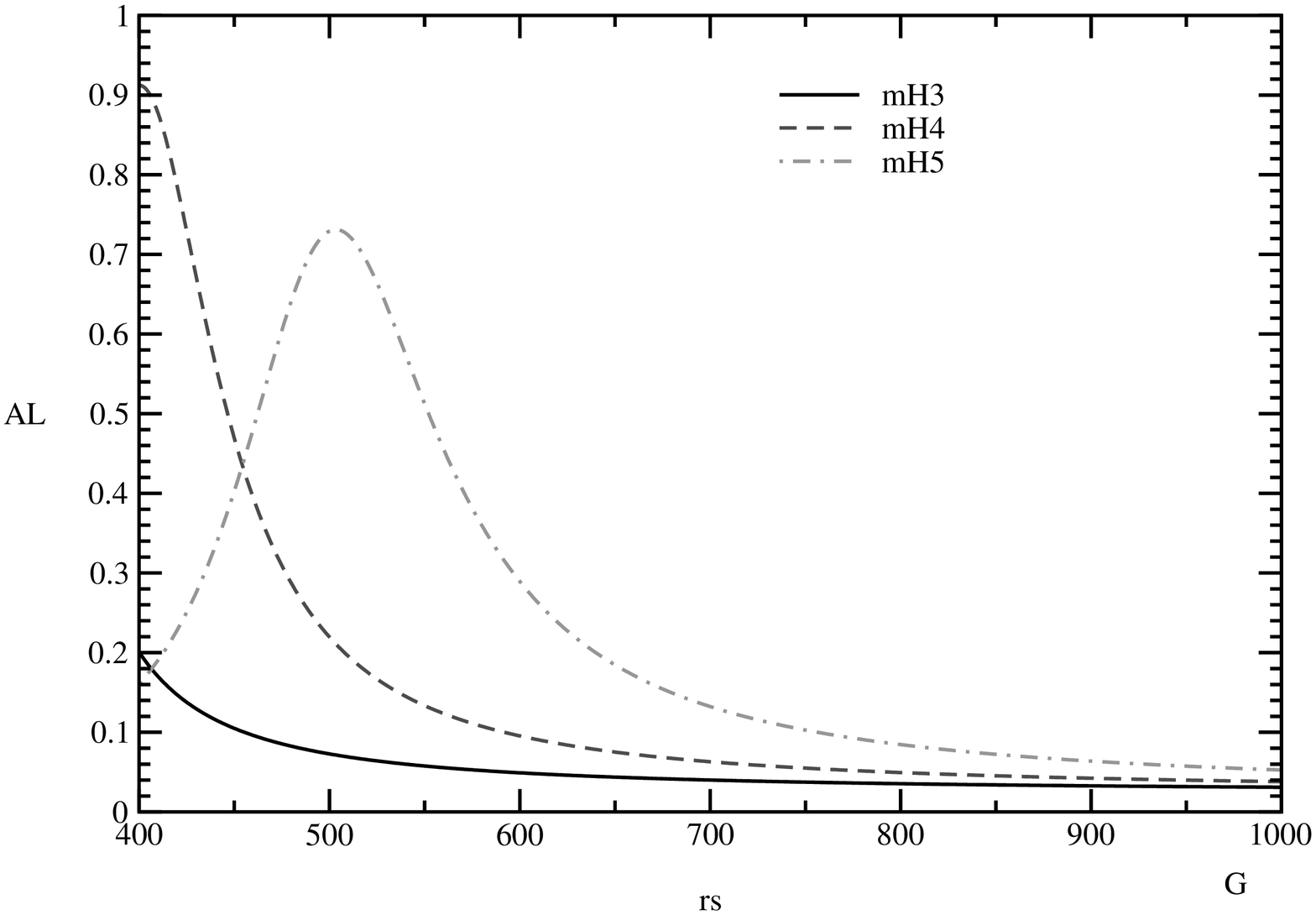}}
\caption{$\sqrt{s}$ dependence of $A_L$~~~~~~~~~~~~~~~~~~~~~~~}\label{ALS}
\vspace*{1.5cm}
 \psfrag{AL}{$A_L$}
 \psfrag{mH}{$m_H$}
 \psfrag{G}{GeV}
 \psfrag{rs5}{\begin{footnotesize}$\sqrt{s}=$~500 GeV\end{footnotesize}}
 \psfrag{rs1}{\begin{footnotesize}$\sqrt{s}=$1000 GeV\end{footnotesize}}
\rotatebox{270}{\includegraphics[width=9.5cm,height=10.5cm]{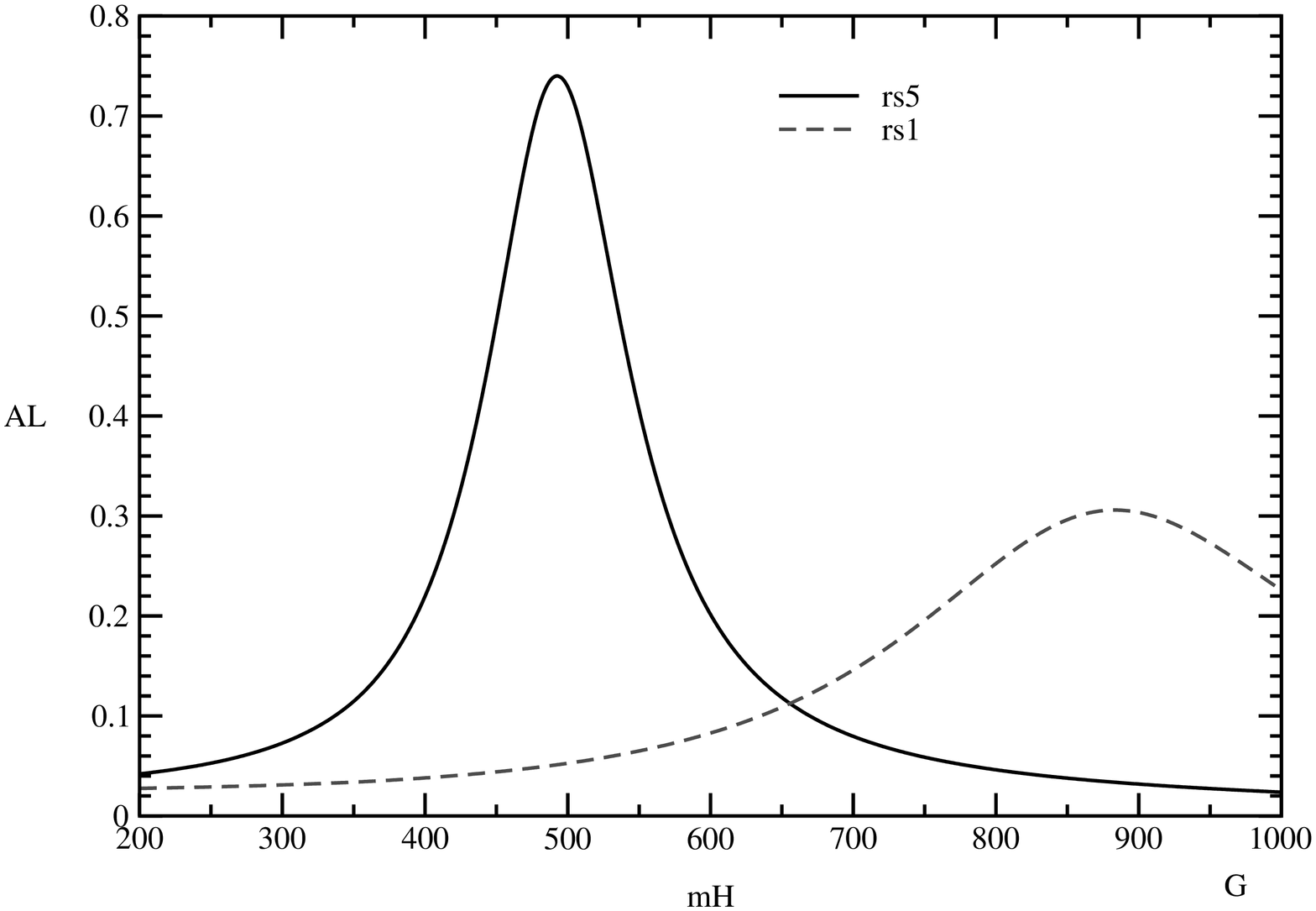}}
\caption{$m_H$ dependence of $A_L$~~~~~~~~~~~~~~~~~~~~~~~}\label{ALMH}
\end{minipage}
\end{figure}

\begin{figure}[htbp]
\hspace*{1.5cm}
\begin{minipage}{14cm}
 \psfrag{AT}{$A_T$}
 \psfrag{rs}{$\sqrt{s}$}
 \psfrag{G}{GeV}
 \psfrag{mH3}{\begin{footnotesize}$m_H=$300 GeV\end{footnotesize}}
 \psfrag{mH4}{\begin{footnotesize}$m_H=$400 GeV\end{footnotesize}}
 \psfrag{mH5}{\begin{footnotesize}$m_H=$500 GeV\end{footnotesize}}
\rotatebox{270}{\includegraphics[width=9.5cm,height=10.5cm]{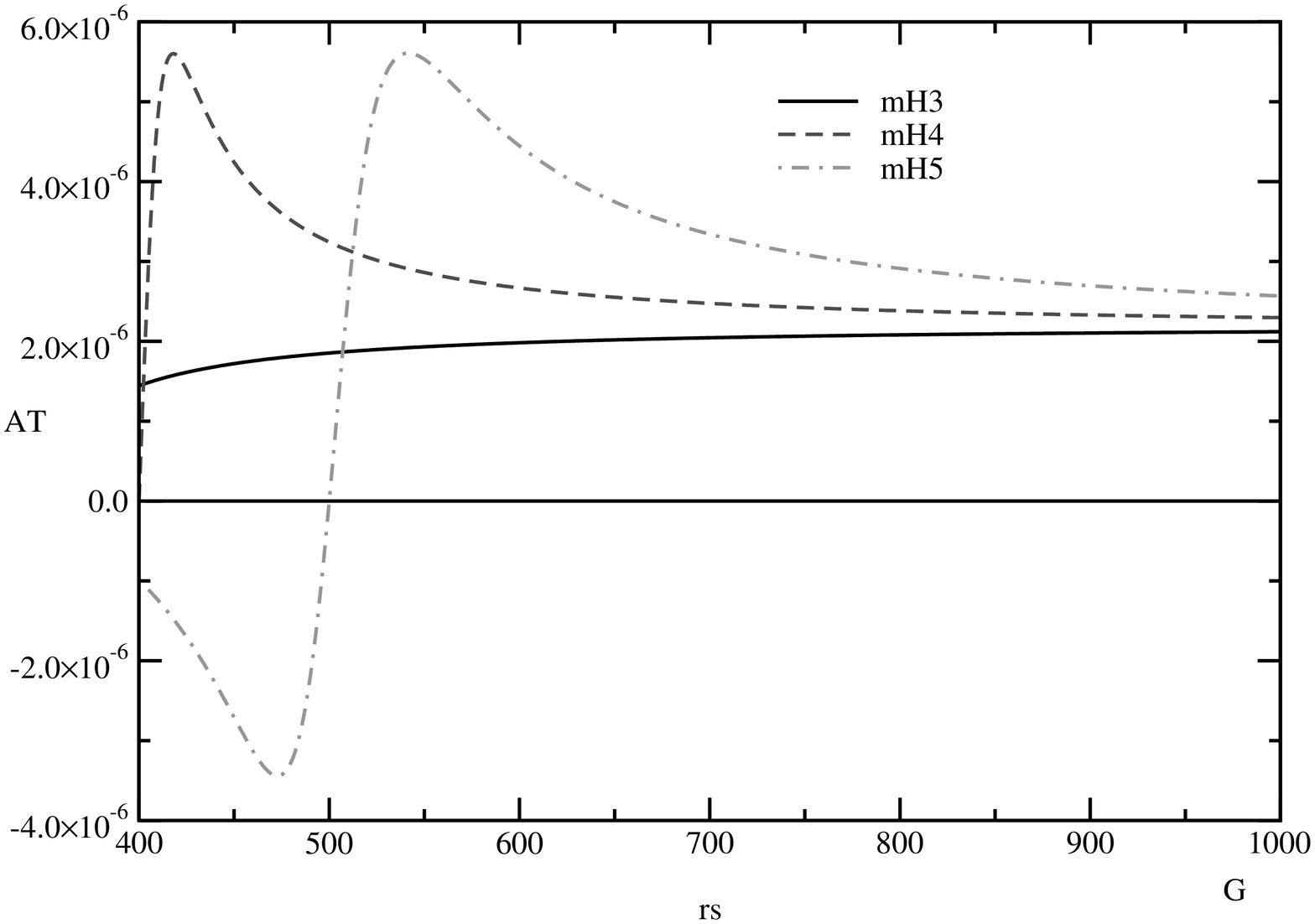}}
\caption{$\sqrt{s}$ dependence of $A_T$~~~~~~~~~~~~~~~~~~~~~~~}\label{ATS}
\vspace*{1.5cm}
 \psfrag{AT}{$A_T$}
 \psfrag{mH}{$m_H$}
 \psfrag{G}{GeV}
 \psfrag{rs5}{\begin{footnotesize}$\sqrt{s}=$~500 GeV\end{footnotesize}}
 \psfrag{rs1}{\begin{footnotesize}$\sqrt{s}=$1000 GeV\end{footnotesize}}
\rotatebox{270}{\includegraphics[width=9.5cm,height=10.5cm]{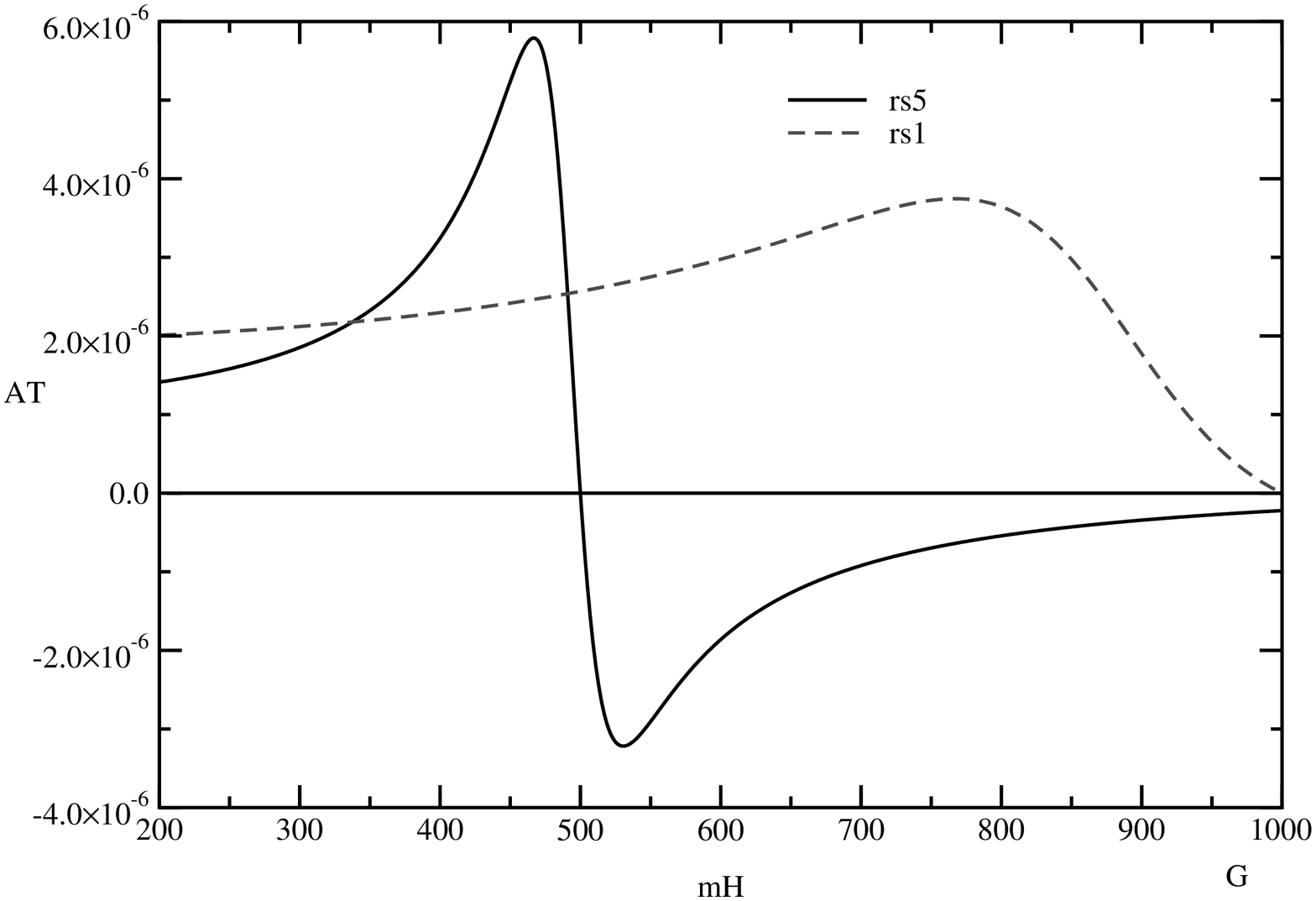}}
\caption{$m_H$ dependence of $A_T$~~~~~~~~~~~~~~~~~~~~~~~}\label{ATMH}
\end{minipage}
\end{figure}

In Figures \ref{ALS} and \ref{ALMH} are presented $A_L$ as functions of
$\sqrt{s}$ and $m_H$, while in Figures \ref{ATS} and \ref{ATMH} are given
$A_T$ in the same way for ${\rm Re}\,a_{t,\mu}={\rm Im}\,a_{t,\mu}
={\rm Re}\,b_{t,\mu}={\rm Im}\,b_{t,\mu}=0.2$ as an example to sketch a
rough feature of these quantities. We find that the absolute value of
$A_L$ could be sizable, but that of $A_T$ is very small. 


It may seem strange that there appears such big difference between
$|A_L|$ and $|A_T|$ though they are both $C\!P$-violating asymmetries computed
with the same anomalous parameters. The reason is in their denominators.
In the case of $A_L$, not only the numerator $\sigma(++)-\sigma(--)$ but
also the denominator receives little contribution from $\gamma/Z$ exchange
terms, while they can contribute to $\sigma(\chi=\pm \pi/2)$ without being
suppressed except in the difference $\sigma(\chi=\pi/2)-\sigma(\chi=-\pi/2)$.
Indeed if we focus on the numerators alone, there is only small difference
between $A_L$ and $A_T$. For $\sqrt{s}=550$ GeV and $m_H=500$ GeV with the same
anomalous couplings as in the figures, e.g.,
\begin{eqnarray}
&&A_L:\ \ \sigma(++)-\sigma(--)=1.6\times 10^{-2}\ {\rm fb}, \\
&&A_T:\ \ \sigma(\chi=\pi/2)-\sigma(\chi=-\pi/2)=4.8\times 10^{-3}\ {\rm 
fb}.
\end{eqnarray}

It is obvious that the peaks in $A_{L,T}$ are all due to the $H$ propagator,
but readers might wonder why $A_T$ changes its sign while $A_L$ not in the
vicinity of $s=M_H^2$. Therefore, it would also be helpful to give a brief
explanation here about those different behaviors of them. As mentioned above,
$\sigma(\chi=\pm \pi/2)$ receive sizable contributions from $\gamma/Z$
exchange terms. This means the interference between the $\gamma/Z$-exchange
and $H$-exchange terms, which is proportional the $H$ propagator, is
important in $A_T$, and its sign changes thereby depending on whether
$s > M_H^2$ or $s < M_H^2$. On the other hand, this is no longer the case
for $A_L$ since the $\gamma/Z$-exchange terms are suppressed in both
$\sigma(++)$ and $\sigma(--)$. Therefore the sign of $A_L$ is determined by
the difference of $\sigma(\pm\pm)$, where the size of the amplitude
${\cal M}_H$ itself is much more crucial. Here we chose $a_\mu = b_\mu$ as
an illustration, which makes $\sigma(++)$ larger than $\sigma(--)$ and leads to
positive $A_L$ as will be understood from eq.(\ref{A++}) on ${\cal M}_H(++)$
and a similar calculation for ${\cal M}_H(--)$. This also tells us that
different parameters, e.g., $a_\mu = -b_\mu$ could make $A_L$ negative.

\noindent
{\bf Detectability of the asymmetry}

Let us study the detectability of $A_L$, that is, the expected
statistical precision in its measurement, which tells us how precisely
we would be able to determine $A_L$. For instance if we take
$\sqrt{s}=m_H=$500 GeV with ${\rm Re}\,a_{t,\mu}
={\rm Re}\,b_{t,\mu}={\rm Im}\,a_{t,\mu}={\rm Im}\,b_{t,\mu}=0.2$, $A_L$
becomes 0.73, while the cross sections are
$\sigma(++)=5.0 \times 10^{-2}$ fb and $\sigma(--)=7.8 \times 10^{-3}$ fb,
leading to $N \simeq 29\epsilon$ events for an integrated luminosity
$L=$ 500 fb$^{-1}$, where we expressed the detection efficiency of $t\bar{t}$
productions as $\epsilon$. They are combined to give the following statistical
uncertainty:
\begin{equation}
\delta A_L = \sqrt{(1-A_L^2)/N} = 0.68/\sqrt{\epsilon L}
= 0.13/\sqrt{\epsilon}.
\end{equation}
Consequently, the expected statistical significance $N_{S\!D}$ is
\begin{equation}
N_{S\!D} \equiv |A_L|/\delta A_L = 5.7 \sqrt{\epsilon}.
\end{equation}
That is, we can confirm $|A_L| \neq 0$ at $5.7 \sqrt{\epsilon}$ level.
For example, $N_{S\!D}= 4.0$ for $\epsilon =0.5$. Here, assuming
$L=$ 500 fb$^{-1}$ may be a bit too optimistic, but we used this value
considering that we aim to find the possibility and limit of the muon
colliders as mentioned in the beginning of this section. It is easy to
transform our numerical results for any other $L$.

We have given an example of $N_{S\!D}$ for $\sqrt{s}=m_H=$500 GeV, but
it is not general, so let us show the results for some other
$\sqrt{s}$ in Table \ref{NSD}, changing also the parameters as ${\rm Re}\,a_{t,\mu}
={\rm Re}\,b_{t,\mu}={\rm Im}\,a_{t,\mu}={\rm Im}\,b_{t,\mu}=0.1,\ 0.2,\ 0.3$.
There we find that we would be able to observe some signal of $C\!P$ violation
as long as we are not too far from the $H$ pole.

\vskip 0.15cm
\begin{table}[thb]
\begin{center}
\begin{tabular}{c|ccc|ccc|ccc}
\lower1.5ex\hbox{$\sqrt{s}$ (GeV)}
&       &  (a)  &          &       &  (b)  &          &       &  (c)  &          \\
& $A_L$ &  $N$  & $N_{S\!D}$ & $A_L$ &  $N$  & $N_{S\!D}$ & $A_L$ &  $N$  & $N_{S\!D}$ \\ \hline
 450 & 0.08  &  \ 7.6 &  0.2  & 0.40  &  11.7 &  1.5  & 0.73  &  25.7 & \ 5.4 \\
 480 & 0.19  &  \ 9.4 &  0.6  & 0.64  &  21.1 &  3.8  & 0.88  &  61.8 &  14.4 \\
 500 & 0.26  &   10.5 &  0.9  & 0.73  &  28.7 &  5.7  & 0.92  &  91.6 &  21.8 \\
 520 & 0.23  &   10.1 &  0.7  & 0.69  &  25.2 &  4.8  & 0.90  &  77.8 &  18.2 \\
 550 & 0.12  &  \ 8.8 &  0.4  & 0.51  &  15.9 &  2.4  & 0.81  &  40.9 & \ 8.9 \\
 600 & 0.05  &  \ 7.7 &  0.1  & 0.29  &  10.4 &  1.0  & 0.63  &  19.8 & \ 3.6 \\ \hline
\end{tabular}
\caption{$N_{S\!D}$ as a function of $\sqrt{s}$ (with $\epsilon=1$ for simplicity) for
${\rm Re}\,a_{t,\mu}={\rm Re}\,b_{t,\mu}={\rm Im}\,a_{t,\mu}={\rm Im}\,b_{t,\mu}=0.1$ (a),
0.2 (b), and 0.3 (c)}\label{NSD}
\end{center}
\end{table}

We used the SM formula for ${\mit\Gamma}_H$ as an appropriate
approximation (${\mit\Gamma}_H=$67.5 GeV for $m_H=$500 GeV \cite{Hwidth}), since we
did not introduce any new light particles that can appear in the final
state of $h$ and $H$ decays. Strictly speaking, however, a new mode like
$H \to hh$ might be possible for $M_h=$ 150 GeV and $M_H=$ 500 GeV.
Instead of re-computing ${\mit\Gamma}_H$ including such new modes, which
demands us to assume a concrete form of those couplings, we give $N_{S\!D}$
for ${\mit\Gamma}_H=$ 80 and 100 GeV with ${\rm Re}\,a_{t,\mu}={\rm Re}\,b_{t,\mu}
={\rm Im}\,a_{t,\mu}={\rm Im}\,b_{t,\mu}=0.2$ in Table \ref{GammaH}, which tells us that
our conclusion would not be affected so much, especially in the off-resonance region.

\vskip 0.15cm
\begin{table}[thb]
\begin{center}
\begin{tabular}{cccc}
$\sqrt{s}$ (GeV) & ${\mit\Gamma}_H={\mit\Gamma}_h(m_H)$ & ${\mit\Gamma}_H=$ 80 
GeV
& ${\mit\Gamma}_H=$ 100 GeV \\ \hline
 450   &  1.5          & 1.4  &  1.3      \\
 500   &  5.7          & 4.4  &  3.2      \\
 550   &  2.4          & 2.3  &  2.1      \\
 600   &  1.0          & 1.0  &  1.0      \\  \hline
\end{tabular}
\caption{$N_{S\!D}$ as a function of $\sqrt{s}$ for ${\mit\Gamma}_H=$ 80 and 
100 GeV, where ${\mit\Gamma}_H={\mit\Gamma}_h(m_H)$ means that ${\mit\Gamma}_H$ was
computed with the SM formula (${\mit\Gamma}_h(m_H)=67.5$ GeV).}\label{GammaH}
\end{center}
\end{table}

\noindent
{\bf Parameter dependence of the asymmetry}

Measuring $A_L$ would be quite interesting, but what $A_L$ receives is
of course one single combination of contributions from all the anomalous
parameters. We are performing a model-independent analysis of possible
new-physics effects, but once we get actual experimental data, our results
are going to be applied for a realistic model building. If $A_L$ does not
depend on some parameters so much, it will be hard to test any models in
which those parameters play a significant role. Therefore it must be
important to see how $A_L$ depends on each parameter. 

\begin{table}[b]
\begin{center}
\begin{tabular}{c|cccccccc}
& ${\rm Re}\,a_t$ & ${\rm Re}\,a_\mu$ & ${\rm Re}\,b_t$ & ${\rm Re}\,b_\mu$ 
& ${\rm Im}\,a_t$ & ${\rm Im}\,a_\mu$ & ${\rm Im}\,b_t$ & ${\rm Im}\,b_\mu$ \\ \hline
0.0  & 4.1 & 4.4 & 4.7 & 2.2 & 4.1 & 4.4 & 4.7 & 2.2 \\
0.1  & 4.7 & 5.3 & 4.9 & 3.8 & 4.7 & 5.3 & 4.9 & 3.8 \\
0.2  & 5.7 & 5.7 & 5.7 & 5.7 & 5.7 & 5.7 & 5.7 & 5.7 \\
0.3  & 7.0 & 5.6 & 7.1 & 7.7 & 7.0 & 5.6 & 7.1 & 7.7 \\ \hline
\end{tabular}
\caption{$N_{S\!D}$ as a function of each parameter for $\sqrt{s}=500$ GeV with the rest
being fixed to be 0.2}\label{T500}
\end{center}
\end{table}

Let us study how $N_{S\!D}$ changes when we vary one parameter
from 0.0 to 0.3. The results are given in Tables \ref{T500} and \ref{T550},
where $N_{S\!D}$ are presented for one of the parameters = 0.0, 0.1, 0.2 and
0.3 with the others being fixed to be 0.2. There we observe that $N_{S\!D}$
receives a contribution from every parameter though there are some differences
among them, which indicates that any model will be testable through measuring
$A_L$.

\begin{table}[t]
\begin{center}
\begin{tabular}{c|cccccccc}
& ${\rm Re}\,a_t$ & ${\rm Re}\,a_\mu$ & ${\rm Re}\,b_t$ & ${\rm Re}\,b_\mu$ 
& ${\rm Im}\,a_t$ & ${\rm Im}\,a_\mu$ & ${\rm Im}\,b_t$ & ${\rm Im}\,b_\mu$ \\ \hline
0.0  & 1.1 & 2.0 & 2.0 & 0.4 & 2.3 & 1.5 & 1.9 & 1.8 \\
0.1  & 1.7 & 2.3 & 2.1 & 1.3 & 2.3 & 1.9 & 2.0 & 2.1 \\
0.2  & 2.4 & 2.4 & 2.4 & 2.4 & 2.4 & 2.4 & 2.4 & 2.4 \\
0.3  & 3.2 & 2.4 & 2.9 & 3.6 & 2.6 & 2.8 & 2.9 & 2.5 \\ \hline
\end{tabular}
\caption{$N_{S\!D}$ as a function of each parameter for $\sqrt{s}=550$ GeV with the rest
being fixed to be 0.2}\label{T550}
\end{center}
\end{table}

\sec{Optimal-observable analysis}

The optimal-observable technique \cite{optimal} is a useful tool
for estimating expected statistical uncertainties in various
coupling measurements. Suppose we have a cross section
\begin{equation}
\frac{d\sigma}{d\phi}(\equiv{\mit\Sigma}(\phi))=\sum_i c_i f_i(\phi),
\label{distribution}
\end{equation}
where $f_i(\phi)$ are known functions of the final-state
variables $\phi$ and $c_i$'s are model-dependent coefficients.
The goal is to determine the $c_i$'s. This can be done by using
appropriate weighting functions $w_i(\phi)$ such that $\int w_i(\phi)
{\mit\Sigma}(\phi)d\phi=c_i$. In general different choices for
$w_i(\phi)$ are possible, but there is a unique choice for which the
resultant statistical error is minimized. Such functions are given by
\begin{equation}
w_i(\phi)=\sum_j X_{ij}f_j(\phi)/{\mit\Sigma}(\phi)\,, \label{X_def}
\end{equation}
where $X_{ij}$ is the inverse matrix of ${\cal M}_{ij}$ which
is defined as
\begin{equation}
{\cal M}_{ij}
\equiv \int {f_i(\phi)f_j(\phi)\over{\mit\Sigma}(\phi)}d\phi\,.
\label{M_def}
\end{equation}
When we use these weighting functions, the statistical uncertainty
of $c_i$ is obtained as
\begin{equation}
\delta c_i=\sqrt{X_{ii}\,\sigma_T/N}\,, \label{delc_i}
\end{equation}
where $\sigma_T\equiv\int (d\sigma/d\phi) d\phi$ and $N$ is the total
number of events.

We study whether we could get more information of the anomalous parameters
via this procedure. Here we focus on the longitudinal beam polarization,
since we found that it is practically impossible to catch any new-physics
signal for the transverse beam polarization even when we could fully use
the total cross sections.

In oder to apply this technique to our analysis, we need to express
the angular distribution of the produced top quark in terms of the
anomalous-coupling parameters like eq.(\ref{distribution}). We have
altogether eight independent
parameters since we assumed the all couplings $a_{t,\mu}$ and $b_{t,\mu}$
to be complex. Although our aim is to perform an analysis as
model-independently as possible, it will be too complicated to treat
them all equally. Therefore we here assume that the size of the imaginary
part of each parameter is much smaller than that of its real part. Since
the imaginary part of parameters (form factors) is often produced
through higher order loop corrections in an underlying theory, this
assumption is not unreasonable. We also drop the terms quartic in
the anomalous parameters.

With this reduced parameter set and assumption, the top-quark angular
distribution should be represented as
\begin{eqnarray}
&&\frac{d}{d\cos\theta}\sigma_{\scriptscriptstyle ++}(\mu\bar{\mu}\to 
t\bar{t})
\nonumber \\
&&\ \ \ \ \ \ \
=f_{\rm SM}(\theta) + c_{aa} f_{aa}(\theta) + c_{ab} f_{ab}(\theta)
+ c_{ba} f_{ba}(\theta) + c_{bb} f_{bb}(\theta),
\end{eqnarray}
where $f_{\rm SM}(\theta)$ expresses the SM contribution,
\begin{eqnarray*}
&&c_{aa} \equiv ({\rm Re}\,a_t)({\rm Re}\,a_{\mu}),\ \ \
  c_{ab} \equiv ({\rm Re}\,a_t)({\rm Re}\,b_{\mu}), \\
&&c_{ba} \equiv ({\rm Re}\,b_t)({\rm Re}\,a_{\mu}),\ \ \
  c_{bb} \equiv ({\rm Re}\,b_t)({\rm Re}\,b_{\mu}),
\end{eqnarray*}
and $f_{ij}(\theta)\:(i,j=a,b)$ are all independent of each other.
If we reverse the signs of $c_{ab}$ and $c_{ba}$, we get
$d\sigma_{\scriptscriptstyle --}(\mu\bar{\mu}\to t\bar{t})$.

Practically, however, $f_{ia}(\theta)$ and $f_{ib}(\theta)$ become
equivalent in the limit of $m_{\mu} \to 0$. We can see this as follows:
In calculating $d\sigma_{\scriptscriptstyle ++}$, the muon-spinor part
becomes
\begin{eqnarray}
&&\bar{v}_+(\mib{p}_{\bar{\mu}})(a_{\mu} + b_{\mu}\gamma_5)u_+(\mib{p}_{\mu})
\simeq
\bar{v}_+(\mib{p}_{\bar{\mu}})(a_{\mu} + b_{\mu}\gamma_5)
\frac{1+\gamma_5}{2} u(\mib{p}_{\mu})   \nonumber \\
&&\phantom{\bar{v}_+(\mib{p}_{\bar{\mu}})(a_{\mu} + b_{\mu}\gamma_5)u_+(\mib{p}_{\mu})}
=(a_{\mu} + b_{\mu})
\bar{v}_+(\mib{p}_{\bar{\mu}})u_+(\mib{p}_{\mu})
\label{A++}
\end{eqnarray}
in the limit. That is, $a_{\mu}$ and $b_{\mu}$ contribute almost equally
to $d\sigma_{\scriptscriptstyle ++}$, which leads to
\[
f_{aa}(\theta) \simeq f_{ab}(\theta),\ \ \ \ \ \
f_{ba}(\theta) \simeq f_{bb}(\theta).
\]
Since we keep $m_{\mu}$ finite, it is in principle possible to perform
an analysis treating all $f_{ij}(\theta)$ as independent functions, but
it is clear that we end up having very poor precision thereby
\cite{GHOW}.
Therefore we neglect their differences from the beginning and start from
\begin{equation}
\frac{d}{d\cos\theta}\sigma_{\scriptscriptstyle ++}(\mu\bar{\mu}\to 
t\bar{t})
\simeq f_1(\theta) + c_a f_2(\theta)
+ c_b f_3(\theta),
\end{equation}
where $c_a \equiv c_{aa} + c_{ab}$, $c_b \equiv c_{ba} + c_{bb}$,
$f_1(\theta) = f_{\rm SM}(\theta)$, $f_2(\theta) = f_{aa}(\theta)
\simeq f_{ba}(\theta)$ and $f_3(\theta) = f_{ba}(\theta) \simeq 
f_{bb}(\theta)$.

Using these functions we obtain the following results as the matrix
(\ref{M_def}) for $\sqrt{s}=550$ GeV and $m_H=500$ GeV:
\begin{eqnarray}
&&
M_{11}= 7.75\cdot 10^{-3},\  M_{12}= 5.80\cdot 10^{-2},\
M_{13}= -2.38\cdot 10^{-3}, \nonumber \\
&&
M_{22}= 4.35\cdot 10^{-1},\  M_{23}= -1.79\cdot 10^{-2},\
M_{33}= 7.37\cdot 10^{-4},
\end{eqnarray}
where we used $f_1(\theta)$ for ${\mit\Sigma}(\phi)$ in eq.(\ref{M_def}).
We then compute the $(2,2)$ and $(3,3)$ elements of the inverse matrix of 
$M$:
\[
X_{22}= 3.68 \cdot 10^6,\ \ \ X_{33}=5.38 \cdot 10^8\,.
\]
This means the expected statistical uncertainty in $c_{a,b}$
measurements are
\begin{equation}
\delta c_a = 1.92 \cdot 10^3 /\sqrt{L},\ \ \ \
\delta c_b = 2.32 \cdot 10^4 /\sqrt{L}\,.
\end{equation}
This tells us that we need $L=3.7\cdot 10^6$ fb$^{-1}$ for achieving
$\delta c_a=1$ and $L=5.4\cdot 10^8$ fb$^{-1}$ for $\delta c_b=1$,
which are both far beyond our reach!

We then assume one of the parameters is determined in some other
experiments in order to look for realistic solutions. First, if $c_a$
was unknown (i.e., if $c_b$ was measured elsewhere), the corresponding
precision becomes $\delta c_a=44.5/\sqrt{L}$, i.e., $\delta c_a=1.99$
for $L=500$ fb$^{-1}$. We give also results for some other $\sqrt{s}$
in Table \ref{deltaca}. Conversely, if $c_b$ is undetermined (i.e.,
only $c_a$ is known), we have
$\delta c_b = 539/\sqrt{L}$, i.e., $\delta c_b=24.1$ for $L=500$ fb$^{-1}$.
Some other results are in Table \ref{deltacb}.

\begin{table}[tb]
\begin{center}
\begin{tabular}{cc}
$\sqrt{s}$ (GeV) & $\delta c_a$ \\ \hline
 400   &  $63.7/\sqrt{L}$    \\
 450   &  $41.1/\sqrt{L}$    \\
 480   &  $44.7/\sqrt{L}$    \\
 520   &  $44.7/\sqrt{L}$    \\
 580   &  $61.4/\sqrt{L}$    \\
 600   &  $75.6/\sqrt{L}$    \\  \hline
\end{tabular}
\caption{Expected precision of $c_a$ determination for
$m_H=500$ GeV}\label{deltaca}
\end{center}
\end{table}

\vskip 0.5cm
\begin{table}[htb]
\begin{center}
\begin{tabular}{cc}
$\sqrt{s}$ (GeV) & $\delta c_b$ \\ \hline
 400   &  $168/\sqrt{L}$    \\
 450   &  $227/\sqrt{L}$    \\
 480   &  $329/\sqrt{L}$    \\
 520   &  $447/\sqrt{L}$    \\
 580   &  $878/\sqrt{L}$    \\
 600   &  $1197/\sqrt{L}$    \\  \hline
\end{tabular}
\caption{Expected precision of $c_b$ determination for
$m_H=500$ GeV}\label{deltacb}
\end{center}
\end{table}

Therefore, if the size of $c_{a}$ is $O(1)$, there is some hope to
catch new-physics signal thereby. On the other hand, $|c_b|$
is required to be at least $O(10)$. Note that it is never unrealistic
to assume $|c_{a,b}|$ to be $O(1) \sim O(10)$ as is known in various
models with two (or multi) Higgs-doublets (see, e.g., \cite{Higgs-hunt}
and the references therein).

What we could know via $A_L$ measurements is only on $C\!P$ violation,
while $c_{a,b}$ are both combinations of $C\!P$-conserving and
$C\!P$-violating parameters. Therefore those two approaches could work
complementarily to each other.

\sec{Summary}

We have carried out a model-independent analysis of possible
non-standard Higgs interactions with $t\bar{t}$ and $\mu\bar{\mu}$
through top-quark pair productions at future muon colliders. As
was pointed out in Refs.\cite{First}--\cite{GGP}, the muon colliders are
quite useful for studying the Higgs sector around the resonance. 
Considering those preceding studies, our main purpose here was to
see if we could also draw any useful information in the off-resonance
region without depending on any specific models.

Starting from
the most general covariant amplitude, we computed two $C\!P$-violating
asymmetries for longitudinal and transverse beam polarizations in
order to see if we could get any signal of new-physics which breaks
$C\!P$ symmetry, and also studied whether we could determine the
non-standard-coupling parameters separately through the
optimal-observable (OO) procedure as a more detailed analysis.

We found that the longitudinal $C\!P$-violating asymmetry $A_L$ would be
sizable, while the transverse asymmetry $A_T$ is too small to be a
meaningful observable. We then estimated the detectability of $A_L$
and showed that we would be able to observe some signal of $C\!P$ violation
as long as we are not too far from the $H$ pole. We also studied there
in some detail how $A_L$ depends on each parameter, and found that we
have no parameter that contribute little, although there are some
differences among the parameters.

On the other hand, more detailed
analyses via the OO procedure seem challenging. However, if we could
reduce the number of unknown parameters with a help of other experiments,
and if the size of the parameters is at least $O(1) \sim O(10)$, we
might be able to get some meaningful information thereby. Readers may
claim that the use of the asymmetry $A_L$ is enough when we have only
one unknown parameter, but this is not necessarily true. What we could
draw from $A_L$ is information on pure $C\!P$ violation, while we could
also know something about $C\!P$-conserving part through an OO analysis. 

In our approach, we need the total cross section of $\mu\bar{\mu} \to
t\bar{t}$ and the angular distribution of the final top quark, for
which we only have to reconstruct the top-quark jet axis. If we further
try to study, e.g., the final lepton distributions in the
top-quark decays, we could get additional information on possible
 anomalous $tbW$ coupling, however in that case we would suffer from
another suppression factor, i.e., the branching ratio of the top-quark
semileptonic decay. Therefore it will be more advantageous to use
the top-quark distribution as a whole when performing off-resonance
analyses at muon colliders.


\vspace{0.6cm}
\centerline{ACKNOWLEDGMENTS}

\vspace{0.3cm}
We are grateful to Federico von der Pahlen for giving us a valuable
comment about the integrated luminosity we used in the text and also
pointing out some typos in the manuscript.
This work is supported in part by the Grant-in-Aid for Scientific
Research No.13135219 and No.16540258 from the Japan
Society for the Promotion of Science, and the Grant-in-Aid
for Young Scientists No. 17740157 from the Ministry of
Education, Culture, Sports, Science and Technology of Japan.
The algebraic calculations using FORM were carried out
on the computer system at Yukawa Institute for Theoretical
Physics (YITP), Kyoto University.

\vspace*{0.8cm}

\end{document}